\documentstyle[psfig]{l-aa}

\begin{document}

\thesaurus{
       (12.03.3;  
        12.12.1;  
        11.03.1)  
          }


\title{ The use of Minimal Spanning Tree to characterize the 2D cluster galaxy
distribution
	\thanks{http://www.astrsp-mrs.fr/www/enacs.html}
      }

\author{C.~Adami \inst{1,2}, A.~Mazure \inst{1}}
\institute{IGRAP, Laboratoire d'Astronomie Spatiale, Marseille, France 
   \and Northwestern University, Chicago, USA }

\offprints{C.~Adami}
\date{Received date; accepted date}

\maketitle
\markboth{The use of Minimal Spanning Tree to characterize the 2D cluster galaxy
distribution}{}

\begin{abstract}

We use the Minimal 
Spanning Tree to characterize the aggregation level of given sets
of points. We test 3 distances based on the histogram of the MST edges 
to discriminate between the distributions. We calibrate the method by 
using artificial sets following Poisson, King or NFW distributions. 
The distance using the mean, the dispersion and the skewness of the  
histogram of MST edges provides the more efficient results.
We apply this distance to a subsample of the ENACS clusters and
we show that the bright galaxies are significantly more 
aggregated than the faint ones. The contamination provided by 
uniformly distributed field galaxies is neglectible. On the other hand, we 
show that the presence of clustered groups on the same cluster line of sight
masked the variation of the distance with the considered magnitude.

\end{abstract}


\section{Introduction}

Clusters of galaxies are the largest (partially) virialized structures in the 
Universe and the cosmological parameters may have a strong influence 
on their profile. The determination of the cluster density profile shape is 
a crucial question.
A high value of $\Omega $ gives for example steeper asymptotic
profiles in the simulations (i.e. Crone et al 1994, Jing et al. 1995). It is
then possible to recover the value of $\Omega $ with the shape of the
clusters. We note however that other studies (see
Navarro et al. 1995, 1996) argue that the dark matter profiles in clusters
deduced from the CDM model are identical whatever the details of the 
model.

Similarly, it is important to know if the cluster profiles exhibit a cusp 
(e.g. Adami et al. 1998). After an analyze of a subsample of the ENACS
clusters, Adami et al (1998) conclude that clusters have a core if we 
consider the
galaxies brighter than b$_j$ = 20. The ENACS clusters (e.g. Katgert et al.
1996 or Mazure et al. 1996) obey the model of a relaxed system
(e.g.\ King 1962). However, the bright galaxies (B$_j\leq $-18.5) are
equally fitted by a profile with core or with cusp. It is then crucial to
know how the shape of the galaxy distribution vary with the magnitude of the
tested galaxies. Adami et al. (1998) have fitted different profiles with different 
shape for the considered magnitudes. This method is very efficient and 
quantitative but also time consuming and complex according to the large 
number of parameters.

We develop here a new way to characterize the variation with magnitude
of the aggregation of the galaxies in clusters, without any profile fitting.
We use the Minimal Spanning Tree (or MST hereafter) which is common
in astronomy to the study of the very large scale structures (e.g. Barrow et 
al. 1985, Bhavsar \& Splinter 1996 or Krzewina \& Saslaw 1996). It is also used 
in physic to study order and disorder of a given set of points (e.g. Dussert et
al. 1986). We use here this last aspect to study the density
profiles of clusters of galaxies by using only a bidimensionnal analysis.
The first part of the article is about the MST theory and the calibration of
the method. We apply the method in the second part to a subsample of 15 very
rich and very regular clusters in order to calibrate the method. The last
part is our conclusions. 

We use H$_0$=100 km.Mpc$^{-1}$.s$^{-1}$ and q$_0$=0.
\section{The MST: theory and calibration}
\label{s-theocalib}

\subsection{theory}
\label{ss-theory}

The MST is a geometrical construction issued from the graph theory: the used
definitions are given in Dussert (1988). Very briefly, it is a tree
joining all the points of a given set, without a loop and with a minimal
length; each point is visited by the tree only 1 time. The main aspect here
is the unicity of such a construction. For a given set of points, there are
more than 1 MST, but the histogram H of the MST edges is unique. This is
fundamental because it is then possible to completely characterize a set of
points with H.

Traditionally, only the two first momenta of H mean m and dispersion 
$\sigma $ are used. These two parameters are efficient for a gaussian 
distribution. To
characterize some non gaussian distributions, we have to use more advanced
momenta like the skewness s and/or the curtosis c. We have tested the use of 
these parameters in the following. Below, we describe the methodology:

\begin{itemize}

\item  The Prim algorithm (1957) is used to construct the MST and 
compute the histogram H of edges.

\item  A point is chosen at random in the set and is the first MST
element.

\item  A point which is the nearest of the MST point, is joined to the MST 
and removed from the set. The first MST edge is between these two points.

\item  We look for the set point which is the nearest of the MST points,
join it to the MST and remove it from the set. The next MST edge is
between this point and its nearest MST point.

\item  We repeat the operation for all the other set points.
\end{itemize}

This algorithm is designed to be the fastest one to have a MST on a given 
set of points. We normalize the lengths of the MST by using the Beardwood 
et al. (1959)
study. A good approximation of the total length of a MST constructed with a
random set of N points in an area S is $\frac{\sqrt{S\times N}}{N-1}$. So,
we divide all the length by this factor (where S is the area of the maximum
rectangle of the point set). We calculate finally the mean m, the dispersion 
$\sigma $, the skewness s and the curtosis c of H.

\subsection{Calibrations of the method}
\label{ss-calib}

We test our algorithm using simulations. We calculate m,$\sigma $,s,c for 
different sets of simulated points. The points are
generated in a 500$\times $500 boxes. We note here that we normalize the
distances and so, the unity of the box size is not important.

In order to characterize the cuspiness degree of the distributions,
we use three kind of 2D density profiles: points randomly distributed
(Poisson distribution), distributed with a centered King profile (flat
profile in the center) and distributed with a centered NFW profile (cusped
profile in the center: Navarro et al. 1995). We note here that the NFW
expression was for a 3D distribution. Applying it for a 2D set of points
generate a more cusped profile compare to the original 3D NFW.\ However, we
will speak of ''NFW profiles'' hereafter. The way we generated sets of
points with a given profile is described in Adami (1998) and is related to
the techniques described in Press et al. (1992). If $\rho $ is the density
and r the radius, we have: 
\[
\rho _{2D\_King}(r)=\frac 1{1+(\frac r{r_c})^2}
\]
and 
\[
\rho _{2D\_NFW}(r)=(\frac 1{\frac r{r_c}(1+\frac r{r_c})^2})^{2/3}
\]
We will call hereafter r$_c$ the characteristic radius of a given profile.
For the King profile, it is the core radius and for the NFW profile it is a
characteristic radius (no core for this profile). We simulate 8 sets of
random distributions: with 10, 25, 60, 125, 250, 500, 750 and 1000
points.

For each set of points with a given profile and a given size, we proceed 100
realizations and so 100 calculations of m,$\sigma $,s,c. From these data, we
are able to compute the mean value and the dispersion of each parameter m, 
$\sigma $ ,s or c.

\subsubsection{Poisson distributions}
\label{sss-unif}

We plot in figure 1 (m,$\sigma $) and (s,c) for a Poisson distribution.

\begin{figure*}
\vbox
{\psfig{file=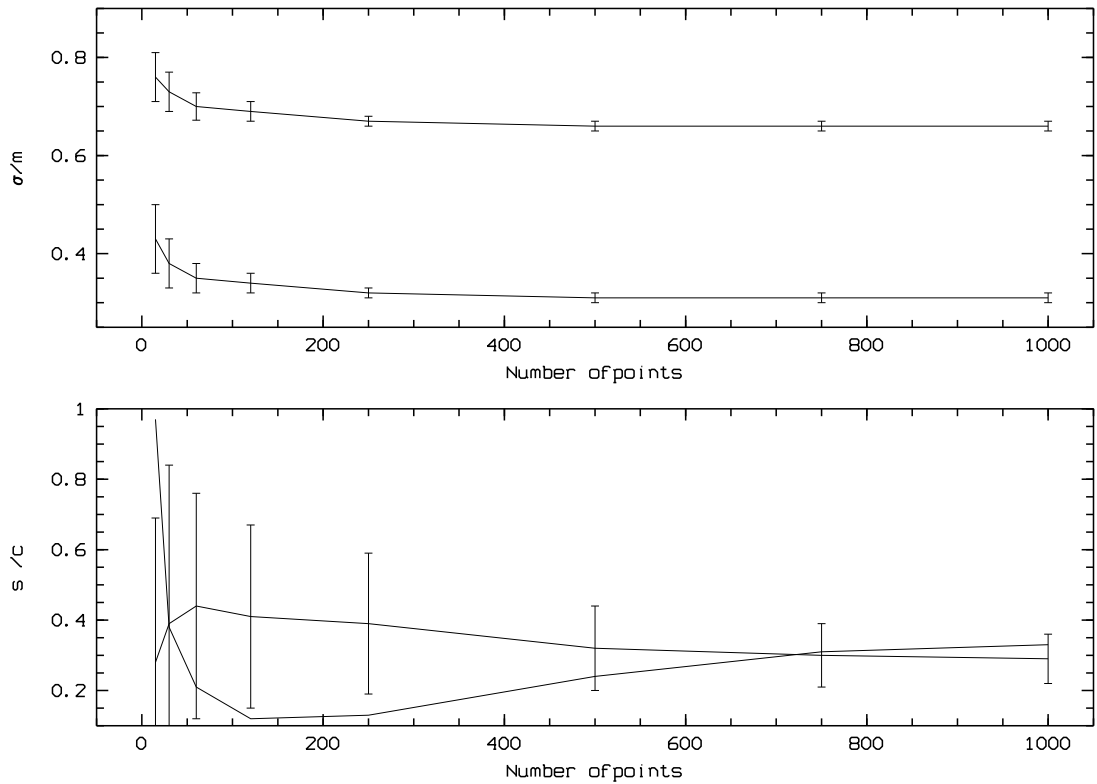,width=17cm,angle=270}}
\caption[]{Variation with the size N of the sample of: up: (m,$\sigma $), m
is the upper line and $\sigma $ is the lower line and down: (s,c), s is the
line with error bars and c is the line without error bar.}
\label{fig1}                                              
\end{figure*}

The parameters m and $\sigma $ are asymptotically equal to 0.66$\pm $0.02
and 0.31$\pm $0.02 in perfect agreement with Dussert (1988). The error bars
are 3\% of the mean value. The final value is reached for a number N of points 
in the simulation greater than 125.

The skewness s is well defined for N$\geq $250 with a final value of 
0.29$\pm $0.14. The errors are greater: about 50\% of the mean value.

The curtosis c is well defined only for N$\geq $750 with very large error
bars ($\sim $100 \% of the mean value). The final value is 0.33.

\subsubsection{King and NFW profiles}
\label{sss-KNFW}

\begin{figure*}
\vbox
{\psfig{file=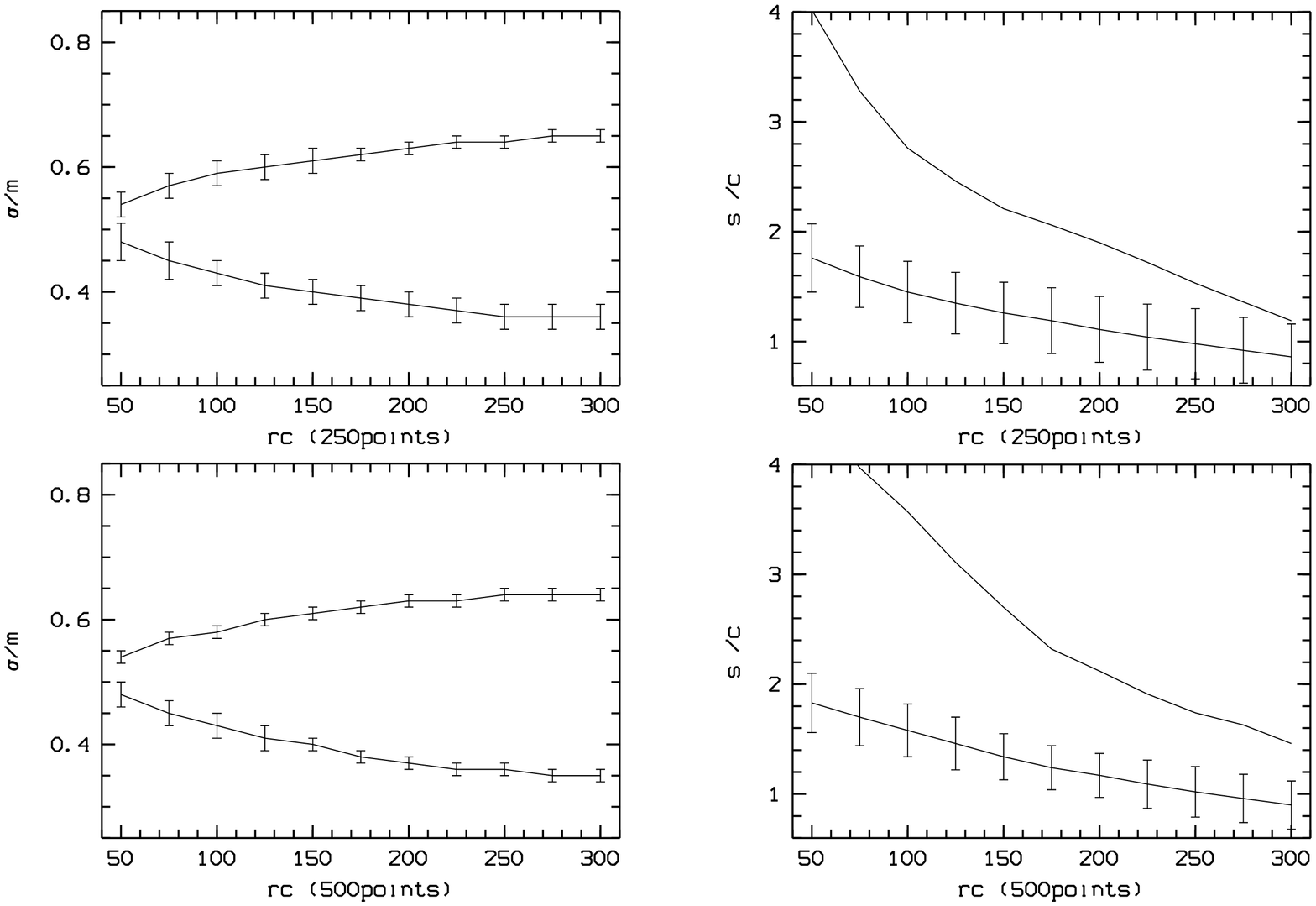,width=17cm,angle=270}}
\caption[]{Variation with the characteristic radius of the King distribution
of: upper left: (m,$\sigma $) and 250 points (the increasing line is m and the
decreasing line is $\sigma $), upper right: (s,c) and 250 points (s is the
line with error bars and c is the line without error bar), lower left: 
(m,$\sigma $) and 500 points (the increasing line is m and the decreasing line 
is $\sigma $), lower right: (s,c) and 500 points (s is the line with error 
bars and c is the line without error bar).}
\label{fig2}
\end{figure*}

We calculate (m,$\sigma $,s,c) for different characteristic radii r$_c$ of
King and NFW profiles. We use r$_c$=50, 75, 100, 125, 150, 175, 200, 225,
250, 275 and 300 kpc. For a given profile and a given characteristic
radius, we simulate 2 sets of points: 250 and 500. We plot the variation
of (m,$\sigma $) and (s,c) with r$_c$ for these 2 sets of points in
figure 2 for the King profiles. The parameters vary significantly with r$_c$
at the 3 $\sigma $ level. The size of the errors are similar to those for
the Poisson case: very small for (m,$\sigma $), median for s and very large
for c. The parameters (m,$\sigma $) are not significantly different from the
Poisson case for large characteristic radii ($\geq $225 kpc). The
skewness is significantly different at the 1 $\sigma $ level from the
Poisson case whatever the characteristic radius. The mean value of the
curtosis is also different, but not significantly because of the large error
bars.

\begin{figure*}
\vbox
{\psfig{file=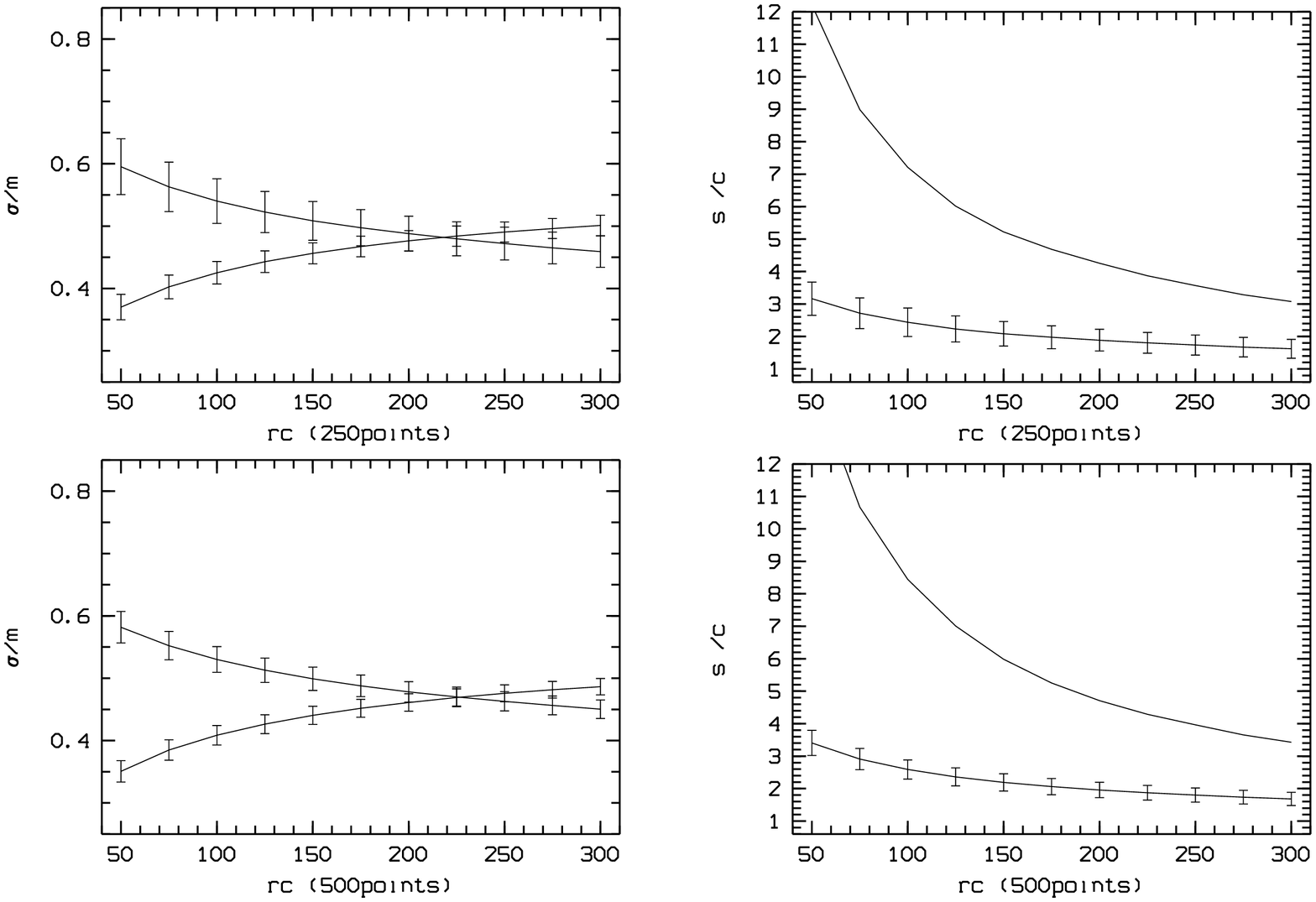,width=17cm,angle=270}}
\caption[]{Variation with the characteristic radius of the NFW distribution
of: upper left: (m,$\sigma $) and 250 points (the increasing line is m and the
decreasing line is $\sigma $), upper right: (s,c) and 250 points (s is the
line with error bars and c is the line without error bar), lower left: 
(m,$\sigma $) and 500 points (the increasing line is m and the decreasing line 
is $\sigma $), lower right: (s,c) and 500 points (s is the line with error 
bars and c is the line without error bar). }
\label{fig3}
\end{figure*}

In figure 3 we plot the variations for the NFW profiles and the trends are
very different. All the parameters (m,$\sigma $,s,c) differ significantly at the
1 $\sigma $ level from the Poisson case. We also notice an important
degeneracy between m and $\sigma $.

\subsubsection{Discrimination between the 3 profiles}
\label{sss-discr}

We want to determine a parameter based on m, $\sigma $, s and c which is able
to discriminate the three profiles. We want to test
the distance in a n dimensional space with n=2, if we use (m,$\sigma $), n=3
if we use (m,$\sigma $,s) and n=4 if we use (m,$\sigma $,s,c). More
generally, the distance in a space of n dimensions between (p$_1$, p$_2$,
....., p$_n$) and (q$_1$, q$_2$, ....., q$_n$) is

\[
\Delta =\sqrt{\sum_{i=1}^n(p_i-q_i)^2} 
\]

The error on a such distance is calculated by derivation:

\[
d\Delta =\frac{\sum_{i=1}^n(p_i-q_i)(dp_i+dq_i)}\Delta 
\]

where $dp_i$ and $dq_i$ are the errors on $p_i$ and $q_i$.

Therefore, we define three distances: $\Delta _{m,\sigma }$, 
$\Delta _{m,\sigma ,s}$, and $\Delta _{m,\sigma ,s,c}$.

We calculate these distances for the Poisson distribution and the King 
and the NFW profiles for 3 different sets of points (50, 125 and 500) and 
for all characteristic radii. We plot (with errors) in figures 4, 5 and 6 these
distances as a variation of the
characteristic radius. We symbolize the nul distance as a solid line and we
plot the error on the determination of the parameters of the Poisson distributions.

\begin{figure*}
\vbox
{\psfig{file=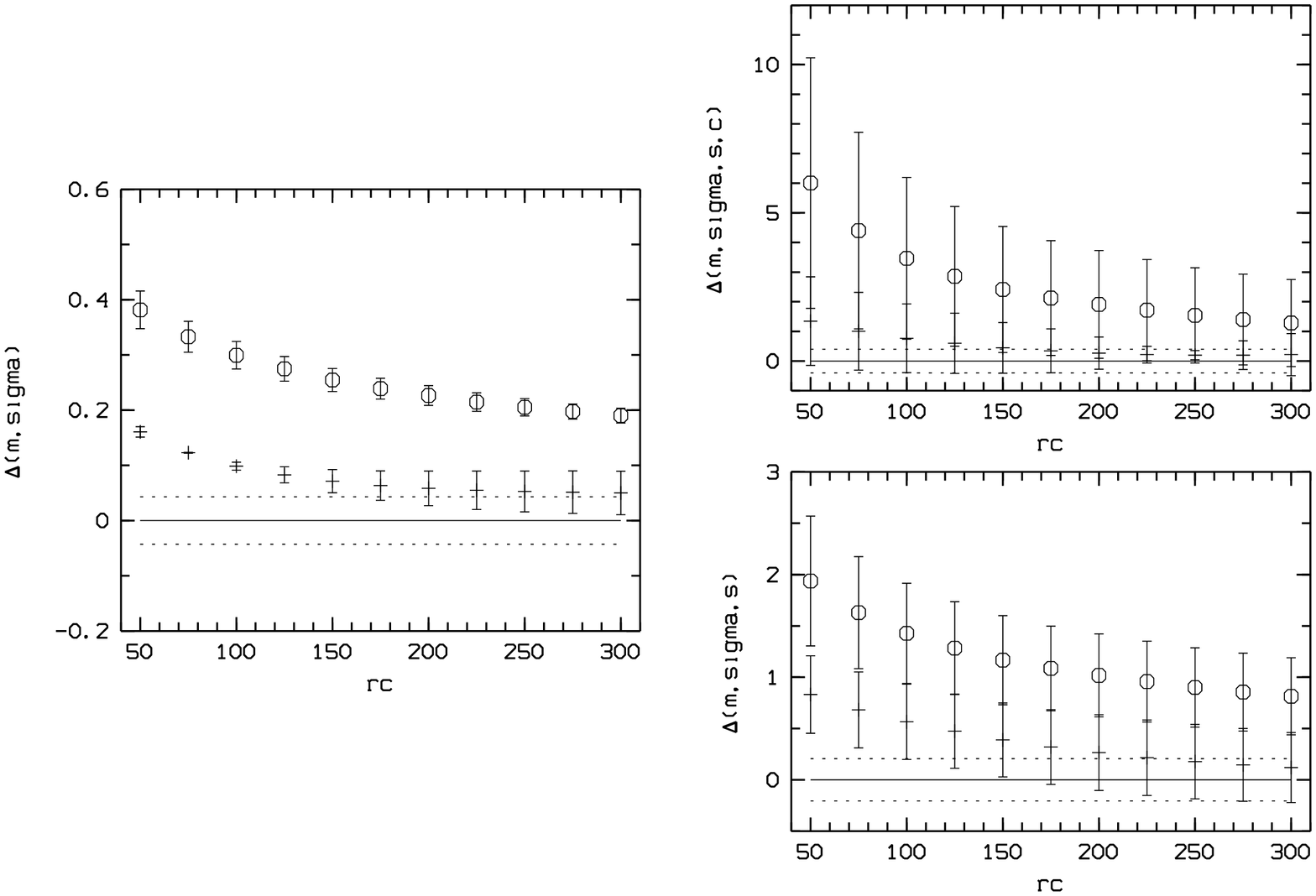,width=17cm,angle=270}}
\caption[]{Variation with the characteristic radius of the 3 tested distances
between the Poisson and the King distribution (crosses) and the Poisson
and the NFW distribution (circles). We have 50 points in the samples.
We plot the error bar on each points and we symbolize the error on the
parameters of the Poisson distribution with the two horizontal dashed lines.
The horizontal solid line symbolizes the nul distance to the Poisson distribution.
The left part of the figure is for the distance $\Delta _{m,\sigma }$, the
lower right part is for $\Delta _{m,\sigma ,s}$ and the upper right part is
for $\Delta _{m,\sigma ,s,c}$.}
\label{fig4}
\end{figure*}

\begin{figure*}
\vbox
{\psfig{file=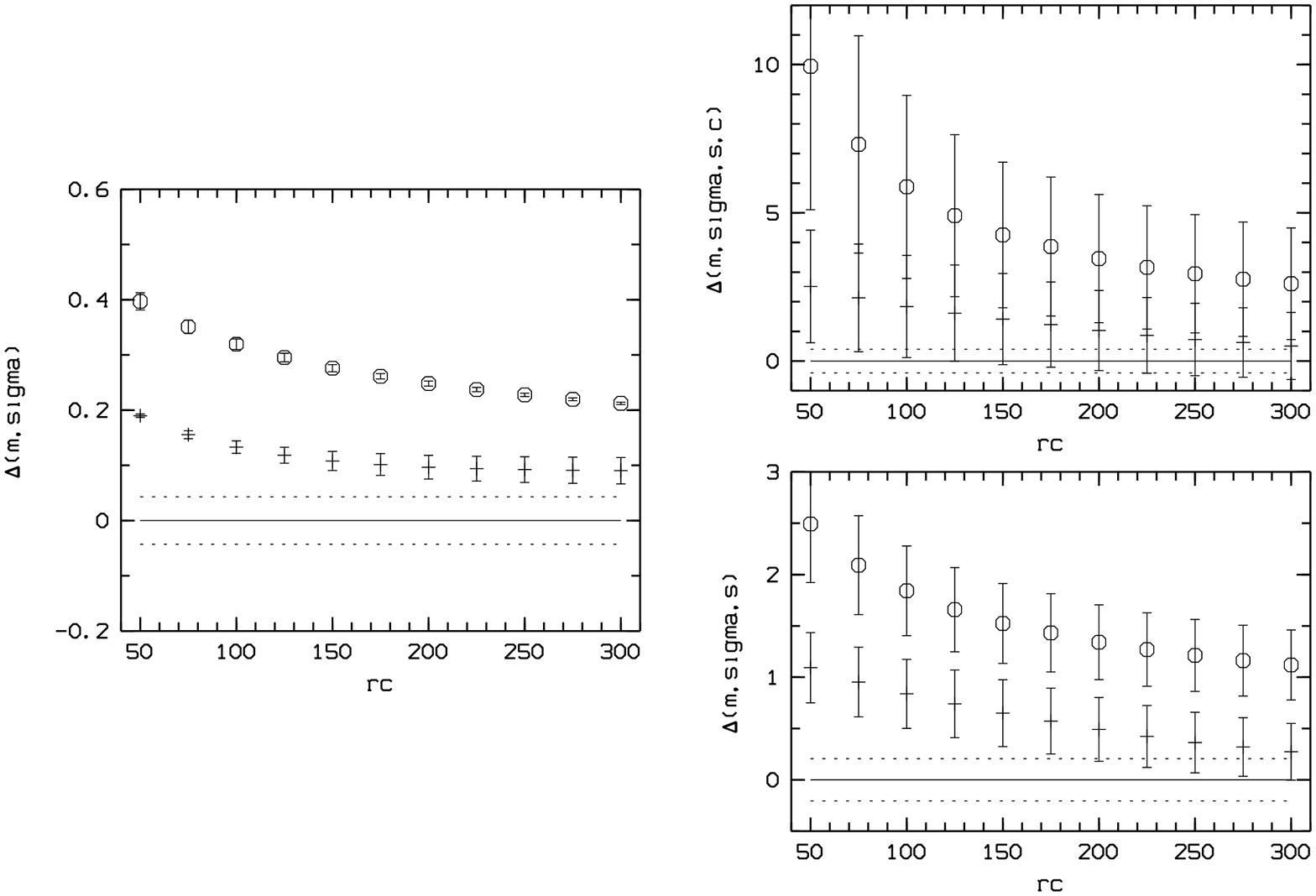,width=17cm,angle=270}}
\caption[]{Variation with the characteristic radius of the 3 tested distances
between the Poisson and the King distribution (crosses) and the Poisson
and the NFW distribution (circles). We have 125 points in the samples.
We plot the error bar on each points and we symbolize the error on the
parameters of the Poisson distribution with the two horizontal dashed lines.
The horizontal solid line symbolizes the nul distance to the Poisson distribution.
The left part of the figure is for the distance $\Delta _{m,\sigma }$, the
lower right part is for $\Delta _{m,\sigma ,s}$ and the upper right part is
for $\Delta _{m,\sigma ,s,c}$.}
\label{fig5}
\end{figure*}

\begin{figure*}
\vbox
{\psfig{file=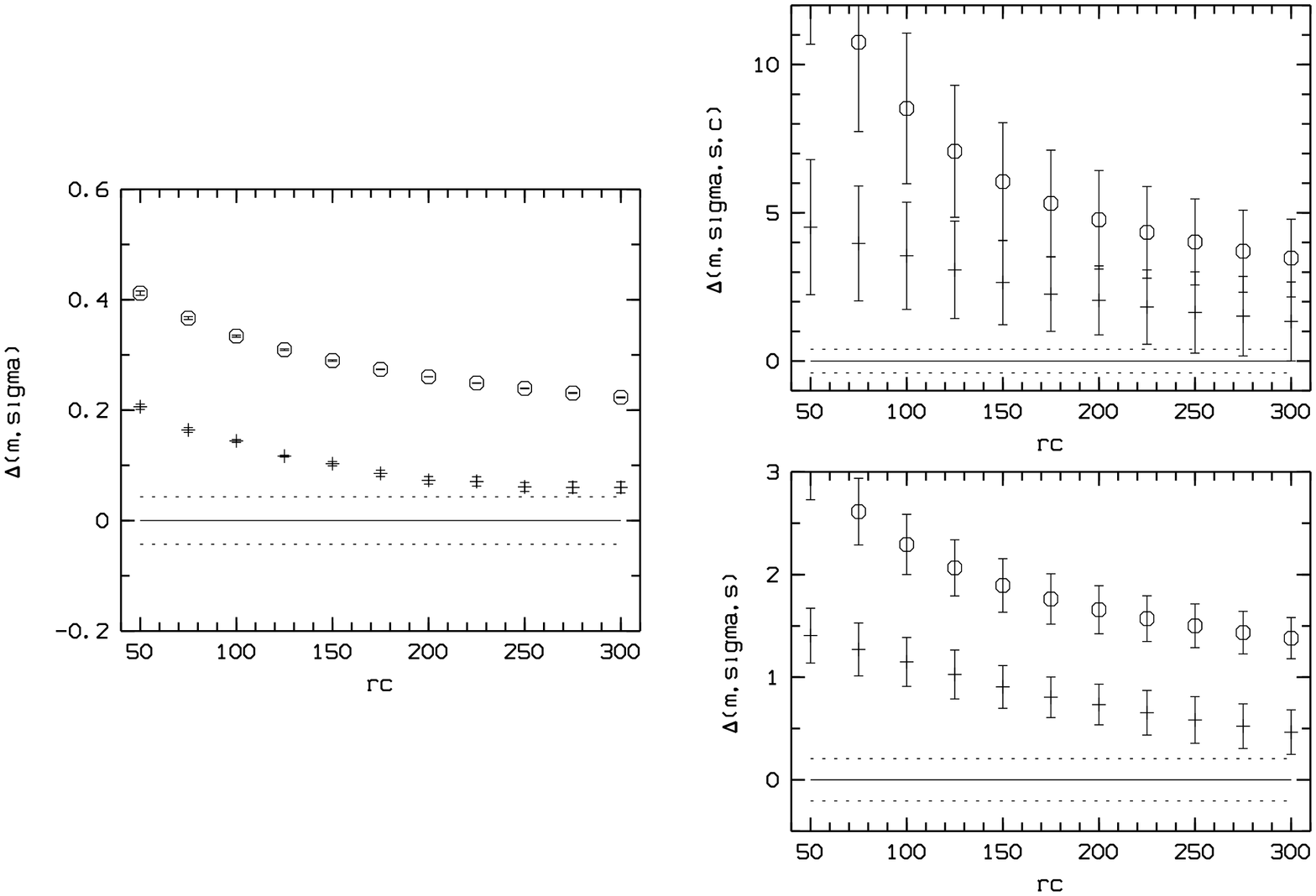,width=17cm,angle=270}}
\caption[]{Variation with the characteristic radius of the 3 tested distances
between the Poisson and the King distribution (crosses) and the Poisson
and the NFW distribution (circles). We have 500 points in the samples.
We plot the error bar on each points and we symbolize the error on the
parameters of the Poisson distribution with the two horizontal dashed lines.
The horizontal solid line symbolizes the nul distance to the Poisson distribution.
The left part of the figure is for the distance $\Delta _{m,\sigma }$, the
lower right part is for $\Delta _{m,\sigma ,s}$ and the upper right part is
for $\Delta _{m,\sigma ,s,c}$.}
\label{fig6}
\end{figure*}

From these figures, we notice the following: 
\begin{itemize}

\item  First of all, the difference 
between the Poisson distribution and King and NFW profiles increases with the 
number of points.

\item  For 50 points, we have much confusion between the King and NFW profiles,
as well as with the
Poisson distribution whatever the distance used. It is impossible to characterize 
the distributions of 50 points.

\item  For 125 or 500 points, $\Delta _{m,\sigma }$ is significant at the 1 
$\sigma $ level between the three profiles. Unfortunately, the values of 
$\Delta _{m,\sigma }$ are very low (more than 60\% lower) compare, for  
example, to the distance between the point (m=0,$\sigma $=0) and the Poisson
distribution. So, the use of $\Delta _{m,\sigma }$ is not straightforward.

\item  For 125 and 500 points $\Delta _{m,\sigma ,s}$ is significant at the
same level, except

between a King profile with r$_c\geq $250 (and 125 objects) and a
Poisson distribution, and

between the King profiles with very low characteristic radii and the NFW
profiles with very high characteristic radii.

The values of $\Delta _{m,\sigma ,s}$ are higher than those of $\Delta
_{m,\sigma }$: 400 \% higher compare to the distance between the point 
(m=0,$\sigma $=0) and the Poisson distribution. The high values and the low 
confusions induced by this distance are able to discriminate efficiently the 
3 profiles. We see a continuous variation of the distance from the Poisson
distributions to the more cusped ones.

\item  For 125 and 500 points, $\Delta _{m,\sigma ,s,c}$ has high values,
but the very large error bars on each parameters induce many confusions
between the King and the NFW profiles (even if the distances with the
Poisson distribution are significant). So, $\Delta _{m,\sigma ,s,c}$ is not the best
distance.
\end{itemize}

We therefore choose $\Delta _{m,\sigma ,s}$ to discriminate between the 
aggregation degree of a set of points. The limiting factor is a number of objects
greater than 125. We note here that, whatever the used distance, we are not
able to discriminate between different characteristic radii for a given
profile, but this is not the goal of this work. The NFW profiles are more
distant to the Poisson distributions than the King ones (whatever
the characteristic radius). According to the cusped and non-cusped shape of
the NFW and King profiles, we can say that the sets of points (with a given
number of points) with a great distance $\Delta _{m,\sigma ,s}$ compare to a
Poisson distribution with the same number of points, are more concentrated
than those ones with a low $\Delta _{m,\sigma ,s}$.

\section{Application to the clusters of galaxies}
\label{s-clusters}

\subsection{A subsample of the ENACS+literature clusters}
\label{ss-enacs}

\begin{table*}
\caption[]{Characteristic parameters of the selected clusters: name,
number of galaxies, slope of the $\Delta _{m,\sigma ,s}$/magnitude
relation, redshift and type of the data (COSMOS/APM).}
\begin{flushleft}
\small
\begin{tabular}{ccccc}
\hline
\noalign{\smallskip}
Cluster name & Number of galaxies & slope of the $\Delta _{m,\sigma ,s}$
magnitude relation & redshift & data \\
\noalign{\smallskip}
\hline
\noalign{\smallskip}
A0168 & 216 & -0.19$\pm $0.05 & 0.045 & COSMOS \\
A0193 & 198 & -0.24$\pm $0.02 & 0.047 & APM \\
A0401 & 424 & -0.30$\pm $0.03 & 0.073 & APM \\
A1069 & 194 & 0 & 0.065 & COSMOS \\
A1367 & 355 & -0.09$\pm $0.02 & 0.023 & APM \\
A2061 & 405 & -0.08$\pm $0.02 & 0.078 & APM \\
A2142 & 314 & +0.20$\pm $0.05 & 0.091 & APM \\
A2670 & 515 & -0.34$\pm $0.03 & 0.075 & APM \\
A2819 & 443 & -0.12$\pm $0.04 & 0.074 & APM \\
A2877 & 586 & -0.19$\pm $0.04 & 0.027 & APM \\
A3112 & 396 & -0.14$\pm $0.04 & 0.075 & COSMOS \\
A3122 & 196 & 0 & 0.068 & COSMOS \\
A3158 & 185 & -0.26$\pm $0.11 & 0.060 & COSMOS \\
A3266 & 299 & 0 & 0.059 & COSMOS \\
A3667 & 754 & -0.60$\pm $0.06 & 0.055 & COSMOS \\
A0401+A1367 & 779 & -0.05$\pm $0.02 & 0.073 + 0.023 & APM \\
\hline
\end{tabular}
\end{flushleft}
\label{t-data}
\end{table*}

The goal of this part is to calibrate the method with a high quality sample 
in order to allow a future more extensive application.
We use a subsample of the regular and richest clusters of galaxies described 
in Adami et al. (1998) and in ENACS IX (1998) to study the variation of the 
galaxy aggregation with magnitude. We have used in these 2 articles COSMOS 
($b_j$ magnitude) and APM surveys (b magnitude). We keep here only the 
richest clusters with more than 180 galaxies 
brighter than $b_j$=20 or b=20 (z$\simeq$0.07 and z$\leq$0.1) in a 5 r$_c$ 
area (At 5 r$_c$ from the center, the surface density is only 1\% of the 
central density if we assume a King profile; Therefore we have the main part 
of the cluster ) and without apparent substructures (15 clusters). We exclude 
finally the clusters with an atypical King core radius (greater than 300 kpc).

We have sorted the galaxies by magnitude (b$_j$ magnitudes). For each cluster, 
we select some sets of 125 consecutive galaxies out of the 180 (or more) 
between the N $^{th}$ and the (N+125)$^{th}$ ranked galaxies. For each of 
those, we calculate the distance $\Delta _{m,\sigma ,s}$ and the error for 
this distance with the corresponding uniform sample. With N=0, 10, 20, 30, 40, 
50, 60 .. etc..., we are able to have many determinations of 
$\Delta _{m,\sigma ,s}$. We note however that these ranges are not 
independent. This allows to compute a variation of the aggregation level with 
the magnitude. We search for a negative slope, characterizing an increasing
aggregation for the bright magnitudes. The selected clusters and the 
characteristic results are listed in table 1. 
                 
The clusters A1069, A3122 and A3266 have not a significant tendency at
the 1 $\sigma $ level: the regression line between $\Delta _{m,\sigma ,s}$
and the magnitude have a slope equal to 0. The cluster A2142 shows a positive
slope (i.e. a decreasing aggregation for the bright magnitudes). The other 
clusters (75 $\%$ of the sample) exhibit a significant decreasing tendency at 
the 1 $\sigma$ level. 

\subsection{Field contamination}
\label{ss-field}

Our simulations do not take into account a possible background contribution.
Such a contamination could reduces the efficiency of the discrimination. We
test here two kind of contamination: a uniform one (uniform density of
background galaxies) and a clustered one (presence of secondary groups on
the same line of sight).

\subsubsection{Uniform contamination}
\label{sss-fieldunif}

In order to test this point, we have selected the cluster A3158 in an area
of 2 Mpc. According to the background level computed in Adami et al. (1998),
the ratio (C hereafter) between the background galaxies and the cluster 
members is 3.7. In this area, the $\Delta _{m,\sigma ,s}$ distance is 
significantly different from 0. We are able to see the structure (Fig.~7).

We increase artificially C by uniformly adding galaxies in
the selected field of view. For each set of added galaxies, we make 100
realizations in order to compute an error for $\Delta _{m,\sigma ,s}.$ We show
in figure 7 the variation of $\Delta _{m,\sigma ,s}$ with C. We can see that 
$\Delta _{m,\sigma ,s}$ is significantly different from 0 (according to the
error bars) for C$\leq $5. For 5$\leq $C$\leq $7, $\Delta _{m,\sigma ,s}$ is
different of 0 in more than 50\% of the realizations. For C$\geq $7, we are
not able to distinguish the cluster structure in more than 50\% of the 100
realizations.

We conclude that we are able to make the difference between the cluster and
the field even if the ratio C is equal to 5, and probably 7. The influence
of a uniform background level is therefore minor.

\begin{figure*}
\vbox
{\psfig{file=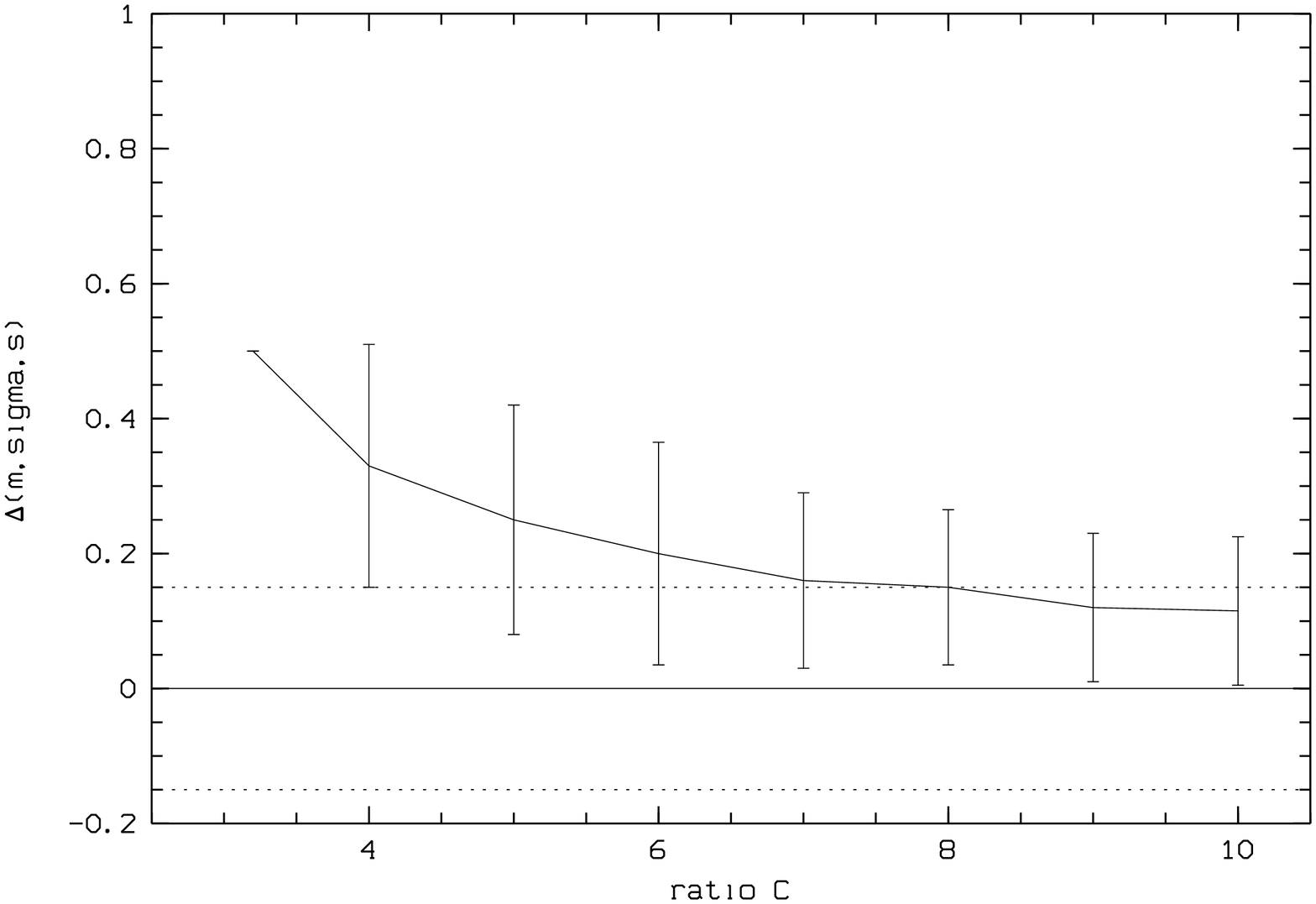,width=17cm,angle=270}}
\caption[]{Variation of $\Delta _{m,\sigma ,s}$ for A3158 with different ratio
C between the number of background and cluster galaxies.}
\label{fig8}
\end{figure*}

\subsubsection{Clustered contamination}
\label{sss-fieldclust}

The other possible contamination is those of secondary groups or clusters on
the same line of sight. To test this effect, we have build a composite
cluster: we have superposed the cluster A0401 (z=0.073) and the cluster
A1367 (z=0.023). The contribution of A1367 will then add a signal of structure 
on the line of sight. We see that we destroy almost all the decrease of 
$\Delta _{m,\sigma ,s}$: the slope of the regression is -0.05$\pm $0.03 (see
Table~1). This kind of contamination can erase the variation of the distance 
$\Delta _{m,\sigma ,s}$ with magnitude. 

\section{Conclusion}
\label{s-conc}

We have constructed a fast method, based on the MST, to characterize the
aggregation level of sets of galaxies. By using the first 3 momenta of the
MST edge length histogram of a given set of galaxies, we are able to
discriminate efficiently between different galaxy distributions compare to 
Poisson ones. The method is not very sensitive to a uniform background
contamination. Even with ratio equal of 5 between field galaxies and cluster
members, we discriminate the structure. The implications of this first
results could be important, in particular to detect possible distant clusters
(or structures) in deep photometric survey. The method also shows that the
bright galaxies clusters are more aggregated than the faint ones in 75 $\%$
of our selected sample. This result is very coherent with Adami et al.
(1998) who show that the bright galaxies in clusters are more aggregated
than the faint ones.

\begin{acknowledgements}
                          
C.A. thanks V. Buat and H. Moulinec for useful discussions and B. Nichol for 
reading an earlier draft of this paper. C.A. and A.M. thanks J.D. Barrow for
useful discussions.

\end{acknowledgements}

\vfill


\begin{thebibliography}{}

\bibitem{}  Adami C., 1998 PhD thesis

\bibitem{}  Adami C., Mazure A., Katgert P., Biviano A., 1998, A\&A sous
presse

\bibitem{}  Barrow J.D., Bhavsar S.D., Sonoda D.H., 1985, MNRAS 216, 17   

\bibitem{}  Beardwood J., Halton J.H., Hammersley J.M., 1959, Camb. Phylos.
Soc. Proc. 55, 299

\bibitem{}  Bhavsar S.P., Splinter R.J., 1996, MNRAS 282, 1461

\bibitem{}  Crone M.M., Evrard A.E., Richstone D.O., 1994, ApJ 434, 402

\bibitem{}  Dussert C., 1988 PhD thesis

\bibitem{}  Dussert C., Rasigni G., Rasigni M., Palmari J., Llebaria A.,
1986, Physical Review B 34, 3528

\bibitem{}  ENACS IX, 1998, A$\&$A in preparation

\bibitem{}  Jing Y.P., Mo H.J., Borner G., Fang L.Z., 1995, MNRAS 276, 417

\bibitem{}  Katgert P., Mazure A., Perea J., et al., 1996, A\&A 310, 8

\bibitem{}  King I.R., 1962, AJ 67, 471

\bibitem{}  Krzewina L.G., Saslaw W.C., 1996, MNRAS 278, 869

\bibitem{}  Mazure A., Katgert P., den Hartog R., et al., 1996, A\&A 310, 31

\bibitem{}  Navarro J.F., Frenk C.S., White S.D.M., 1996, MNRAS 462, 563

\bibitem{}  Navarro J.F., Frenk C.S., White S.D.M., 1995, MNRAS 275, 720

\bibitem{}  Press W., Teutolsky S., Vetterling W., Flannery B., 1992
''Numerical Recipes'' Cambridge University Press

\bibitem{}  Prim R.C., 1957, Bell. Syst. Tech. J. 36, 1389

\end{thebibliography}
\end{document}